%% file: kimreview.tex
\documentclass[11pt, hyphens]{kimreview}
\input{config}
\usepackage{graphicx}
\usepackage{subcaption}
\begin{document}


\title{Event-Chain Monte Carlo: The global-balance breakthrough}


\authorOne{E.A.J.F. (Frank) Peters}
\affiliationOne{Department of Chemical Engineering and Chemistry, Eindhoven University of Technology, The Netherlands}


\publishyear{2026}
\volumenumber{4}
\articlenumber{02}
\submitdate{January 9, 2026}
\publishdate{February 6, 2026}
\doiindex{10.25950/a25ce9a6}
\doilink{10.25950/a25ce9a6}

\paperReviewed
{Event-chain Monte Carlo algorithms for hard-sphere systems}
{E.~P.\ Bernard, W.\ Krauth, D.~B. Wilson}
{\href{https://doi.org/10.1103/PhysRevE.80.056704}{Phys. Rev. E, 80:056704, 2009}}

\maketitle



\begin{abstract}
The seminal 2009 paper by Bernard, Krauth, and Wilson marked a paradigm shift in Monte Carlo sampling. By abandoning the restrictive condition of detailed balance in favor of the more fundamental principle of global balance, they introduced the Event-Chain Monte Carlo (ECMC) algorithm, which achieves rejection-free, deterministic sampling for hard spheres. This breakthrough demonstrated that persistent, directional dynamics could dramatically accelerate equilibration in dense particle systems. In this commentary, we review this foundational work and elucidate its underlying mechanism using the broader Event-Driven Monte Carlo (EDMC) framework developed in subsequent years. We show how the original hard-sphere concept naturally generalizes to continuous potentials and modern lifted Markov chain formalisms, transforming a surprising specific result into a powerful general class of sampling algorithms.
\end{abstract}

\medskip

\section{Introduction: From disbelief to understanding}

The idea introduced in the original Event-Chain Monte Carlo (ECMC) paper by Bernard, Krauth, and Wilson~\cite{Bernard2009} is one of those insights that feels both surprising and inevitable.
It seems almost paradoxical: a Monte Carlo move that never rejects, moves particles deterministically, but samples the correct equilibrium distribution.
When I first read the paper, my immediate thought was: This cannot possibly work.
But after following the reasoning, the elegance of the approach becomes striking.
It is so simple, so physically intuitive, that one wonders: Why was this not discovered decades ago?

\begin{figure}
    \centering
    \begin{subfigure}[t]{0.495\linewidth}
        \centering
        \includegraphics[width=0.95\linewidth]{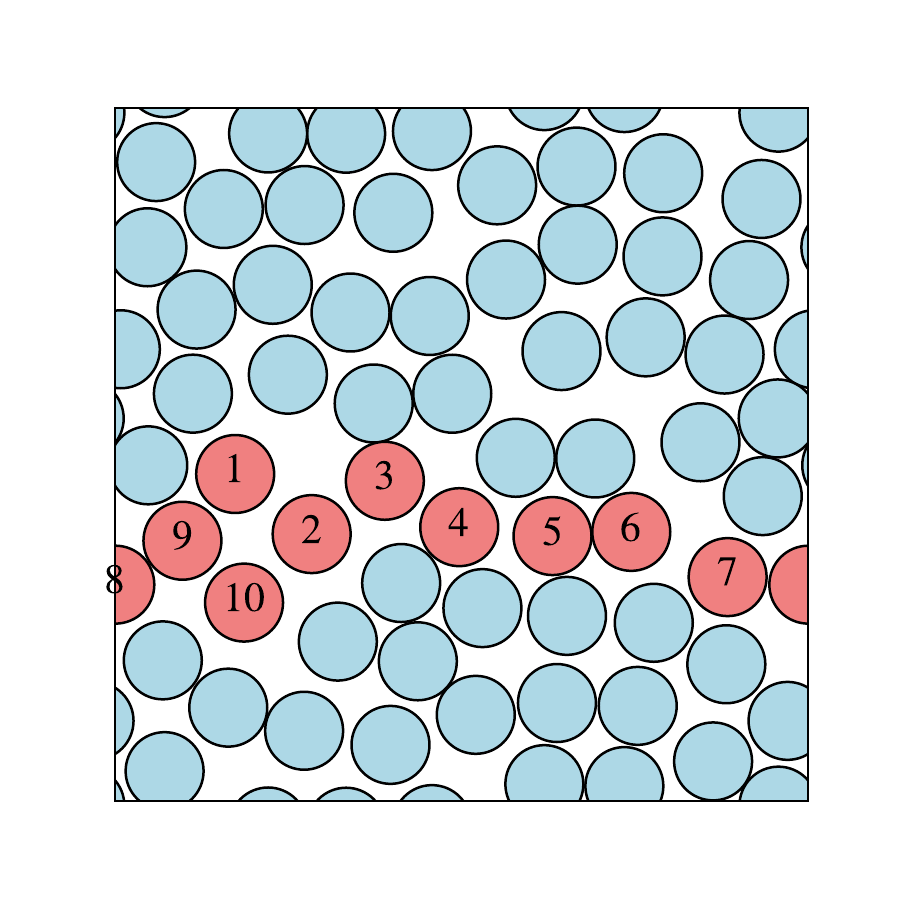}
        \caption{before displacement}\label{fig:event_chain_before}
    \end{subfigure}
    \hfill
    \begin{subfigure}[t]{0.495\linewidth}
        \centering
        \includegraphics[width=0.95\linewidth]{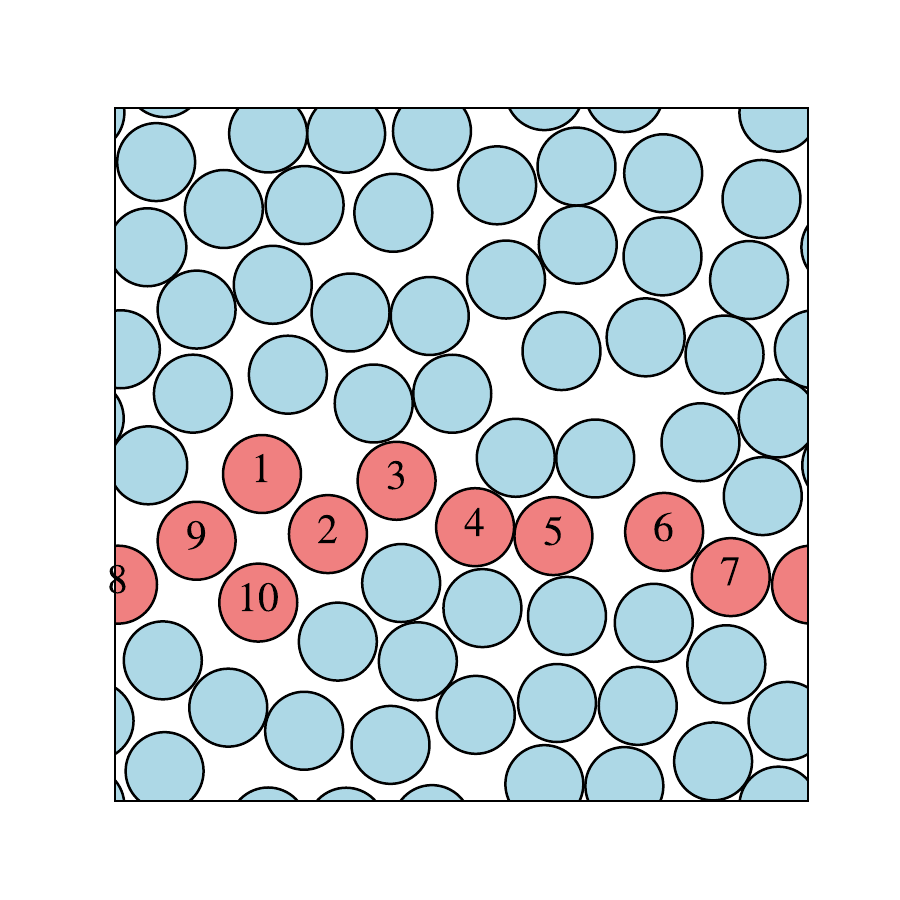}
        \caption{after displacement}\label{fig:event_chain_after}
    \end{subfigure}
    \caption{An event-chain move. The particle $1$ is picked to move to the right. When it collides with particle $2$, particle $1$ halts and $2$ takes over etc.,\ forming an \emph{event} chain. The motion stops of the sum of displacements add up to a preset value, $l$, which was one fifth of the box size in this case. Plot (a) shows the initial positions of particles involved in the chain, and (b) the final positions.}\label{fig:event_chain}
\end{figure}

In essence, ECMC replaces the random, stop-start behavior of Metropolis sampling with continuous, deterministic motion. As illustrated in Fig.~\ref{fig:event_chain}, a single move consists of picking a particle (particle 1) and displacing it in a fixed direction (here to the right) until it collides with another particle (particle 2). At this \emph{event}, the activity is transferred: particle 1 stops, and particle 2 immediately begins moving in the same direction until it collides with a third particle, and so on. This creates a \emph{chain} of displacements.
The entire move stops when the total displacement of all particles in the chain sums to a pre-defined chain length, $l$. Figure~\ref{fig:event_chain_before} shows the initial positions of the particles that will be part of the chain, and Fig.~\ref{fig:event_chain_after} shows their final positions after the chain has completed.

This process is rejection-free. For a given direction and chain length, the move is deterministic. A chain run in the reverse direction would perfectly restore the original configuration. The groundbreaking insight of Bernard et al.\cite{Bernard2009} was to break detailed balance. By only ever moving particles in ``positive'' directions (e.g., $+x$ or $+y$), they create a persistent probability flow. This clearly violates detailed balance. However, as the authors showed, the algorithm still preserves the uniform equilibrium distribution of hard-sphere configurations. Every configuration has as many incoming chains as outgoing chains, thus satisfying the \emph{global balance} condition. In practical terms, ECMC turns the diffusive random walk of Metropolis sampling into a nearly ballistic exploration of configuration space, leading to much faster equilibration.

In this commentary, I aim to provide a didactic exposition of the theoretical and practical underpinnings of this breakthrough. While the physical intuition behind the hard-sphere event chain is easily grasped, the underlying principle generalizes to arbitrary continuous potentials, offering a unified perspective on rejection-free sampling. I have chosen to present a relatively technical derivation, starting from basic probability flows and moving toward the continuous-time limit. This allows us to see how the original ECMC algorithm \cite{Bernard2009} and its subsequent generalizations to Event-Driven Monte Carlo (EDMC) \cite{Peters2012} are unified under the rigorous framework of infinitesimal steps \cite{Michel2014}. My goal is to guide the reader from the initial surprise of the 2009 paper to a concrete understanding of how universal global-balance methods can be constructed.

\section{Global balance}

The surprise of ECMC stems from its departure from the standard recipe for MCMC algorithms, the Metropolis algorithm. Conventional MCMC typically enforces the detailed balance condition, which requires that at equilibrium the probability flow between any two states $i$ and $j$ is exactly balanced. This is a strong, local condition that makes it easy to verify that a given target distribution $\pi$ is stationary for the Markov chain. However, detailed balance is only a sufficient condition. The fundamental requirement for stationarity is the weaker condition of \emph{global balance}.

To make this precise, it is useful to recall the basic structure of a Markov-chain Monte Carlo simulation. The transition from the current state $x_i$ to a new state $x_j$ is drawn from a transition probability $T(x_i \to x_j)$, collected in the transition matrix $T_{ij} \equiv T(x_i \to x_j)$. By construction,
\begin{equation}
p_{n+1}(x_j) = \sum_i p_{n}(x_i) \, T(x_i \to x_j),
\qquad
\sum_j T(x_i \to x_j) = 1,
\label{eq:mcmc}
\end{equation}
where $p_n(x_i)$ is the probability to be in state $x_i$ after $n$ steps. The second relation expresses that the probabilities over all possible next states sum to one. Using this normalization, Eq.~\eqref{eq:mcmc} can be rewritten as
\begin{equation}
p_{n+1}(x_j) - p_{n}(x_j) = \sum_i \left[ p_{n}(x_i) \, T(x_i \to x_j) - p_{n}(x_j) \, T(x_j \to x_i) \right],
\label{eq:master}
\end{equation}
which is the discrete-time master equation for the Markov chain.

The transition matrix $T$ is a (row-)stochastic matrix: all entries are non-negative and each row sums to one. The Perron-Frobenius theorem implies that, if $T$ is irreducible, there exists a unique equilibrium distribution $\pi(x_i)$ to which the chain converges, i.e., $p_n(x_i) \rightarrow \pi(x_i)$ for $n\to\infty$~\cite{Manousiouthakis1999}. Irreducibility means that in a finite number of steps, say $k$, any state can be reached from any other state with non-zero probability; mathematically, all entries of $T^k$ are strictly positive. In the stationary state, Eq.~\eqref{eq:mcmc} reduces to
\begin{equation}
\pi(x_j) = \sum_i \pi(x_i) \,  T(x_i \to x_j),
\label{eq:gobal_balance}
\end{equation}
which is precisely the condition of \emph{global balance}: the total probability flowing into state $x_j$ equals the stationary probability of that state. Using the master-equation form \eqref{eq:master}, the same condition can be written as
\begin{equation}
\sum_i \pi(x_i) \, T(x_i \to x_j) = \sum_i \pi(x_j) \, T(x_j \to x_i),
\label{eq:master_global_balance}
\end{equation}
expressing that, at equilibrium, the net probability current into each state vanishes.

In statistical mechanics, one is interested in sampling equilibrium distributions such as the Boltzmann distribution,
\begin{equation}
\pi(x) = \frac{1}{Z} \exp\left( -\beta \, U(x) \right),
\label{eq:Boltzmann_distribution}
\end{equation}
where $U(x)$ is the potential energy and $Z$ the normalization (partition function). For constructing the transition probabilities $T(x_i \to x_j)$, knowledge of the normalization constant $Z$, since it cancels from all global-balance expressions.
The practical question is then: given $U(x)$, how should one construct and sample a transition kernel $T(x_i \to x_j)$ such that the resulting Markov chain is irreducible, obeys global balance with respect to \eqref{eq:Boltzmann_distribution}, and does so as efficiently as possible? Detailed-balance schemes such as Metropolis provide one answer; ECMC and related global-balance algorithms provide another.

\subsection{Probability flows}

To sample a known equilibrium distribution by means of a Markov-chain Monte Carlo method, one must enforce either the global balance condition, Eq.~\eqref{eq:gobal_balance}, or the equivalent master-equation form, Eq.~\eqref{eq:master_global_balance}.

A simple way to satisfy the stationary master equation, Eq.~\eqref{eq:master_global_balance}, is to require that every term in the sum vanishes individually. This yields the detailed-balance condition
\begin{equation}
\pi(x_i)\, T(x_i \to x_j) = \pi(x_j)\, T(x_j \to x_i),
\label{eq:detailed-balance}
\end{equation}
which is sufficient but not necessary for global balance.

For constructing Monte Carlo algorithms that sample $\pi(x)$ at equilibrium it is often useful to express global balance in terms of equilibrium \emph{probability flows}:
\begin{equation}
\pi(x_i; x_j) \equiv \pi(x_i)\, T(x_i \to x_j).
\label{eq:probability_flow}
\end{equation}
With this definition, global balance, Eq.~\eqref{eq:gobal_balance}, and the normalization of the transition probabilities can be written compactly as
\begin{equation}
\pi(x_j) = \sum_i \pi(x_i; x_j),
\qquad
\pi(x_i) = \sum_j \pi(x_i; x_j).
\label{eq:probability_flow_conditions}
\end{equation}
The quantity $\pi(x_i; x_j)$ is the two-point equilibrium probability of being in state $x_i$ and transitioning to $x_j$ in one step.
Equation~\eqref{eq:probability_flow_conditions} simply expresses that summing this two-point distribution over either of the two states yields the one-point stationary distribution.

\begin{figure}
\centering
\includegraphics[]{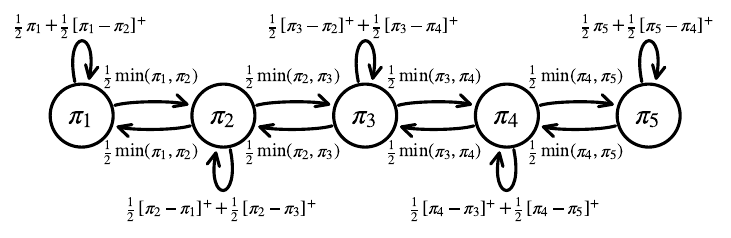}
\caption{The Metropolis scheme in 1D. Incoming and outgoing equilibrium probability flows add up to the stationary probability $\pi_i$. Detailed balance holds because the probability flows between any two nodes are symmetric.}
\label{fig:metropolis}
\end{figure}

In these terms, detailed balance [Eq.~\eqref{eq:detailed-balance}] becomes the symmetry condition: $\pi(x_i; x_j) = \pi(x_j; x_i)$,
which in mathematical literature is often described as a \emph{reversible} Markov chain.
In molecular simulation this terminology is best avoided, since “reversibility’’ is already used in the dynamical sense of Hamiltonian systems; here we continue to refer simply to detailed balance.

One way to construct a Markov chain that satisfies Eq.~\eqref{eq:probability_flow_conditions} is to require
\begin{align}
    \sum_{i\ne j} \pi(x_i; x_j) &= \sum_{i\ne j} \pi(x_j; x_i) \le \pi(x_i),
    \label{eq:condition_minimum} \\
    \pi(x_i; x_i) &= \pi(x_i) - \sum_{j\ne i} \pi(x_i; x_j)
                  = \pi(x_i) - \sum_{j\ne i} \pi(x_j; x_i).
\end{align}
A straightforward way to achieve this is to introduce a symmetric trial kernel,
$T_\mathrm{trial}(x_i \to x_j) = T_\mathrm{trial}(x_j \to x_i)$, and define
\begin{equation}
    \pi(x_i; x_j)
    = T_\mathrm{trial}(x_i \to x_j)\,
      \min\!\big(\pi(x_i), \pi(x_j)\big),
    \qquad i \ne j.
\label{eq:metropolis_flow}
\end{equation}
Because the trial kernel is symmetric, the sums over $i$ and $j$ are equal, and the $\min$ ensures they never exceed $\pi(x_i)$ or $\pi(x_j)$.
The corresponding transition probabilities are
\begin{align}
    T(x_i \to x_j)
    &= \frac{\pi(x_i; x_j)}{\pi(x_i)}
     = T_\mathrm{trial}(x_i \to x_j)\,
       \min\!\left(1, \frac{\pi(x_j)}{\pi(x_i)}\right),
       \qquad i\ne j, \\
    T(x_i \to x_i)
    &= 1 - \sum_{j\ne i} T(x_i \to x_j)
     = \sum_j T_\mathrm{trial}(x_i \to x_j)\,
       \left[ 1 - \frac{\pi(x_j)}{\pi(x_i)} \right]^+ .
\label{eq:metropolis_transition}
\end{align}
Here the notation $[a]^+$ denotes $\max(0,a)$.

Equations~\eqref{eq:metropolis_flow}–\eqref{eq:metropolis_transition} describe the standard Metropolis algorithm:
a proposed move to $x_j$ is accepted with probability $\min(1,\pi(x_j)/\pi(x_i))$, otherwise the chain remains at $x_i$, contributing to the self-transition $T(x_i \to x_i)$.
Figure~\ref{fig:metropolis} illustrates the Metropolis flows in one dimension.
Global balance [Eq.~\eqref{eq:probability_flow_conditions}] requires that the sum of all incoming flows and the sum of all outgoing flows each equal $\pi_i$.

The Metropolis algorithm has two well-known shortcomings that global-balance schemes such as ECMC aim to mitigate.
First, the symmetric trial kernel $T_\mathrm{trial}(x_i \to x_j)$ leads to local diffusive dynamics.
Second, in dense systems the acceptance probability can be very small, resulting in many rejections and correspondingly slow sampling.

\subsection{Lifted Metropolis schemes}

\begin{figure}
    \centering
    \includegraphics[]{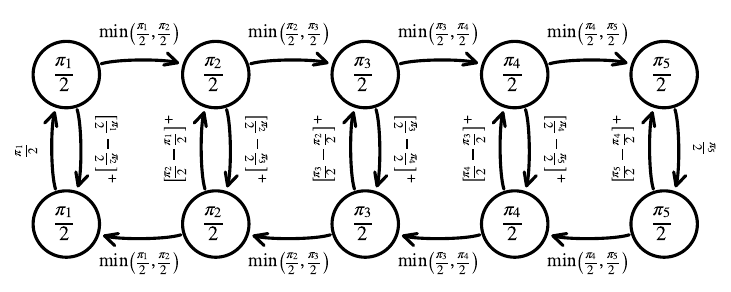}
    \caption{A lifted Markov-chain scheme in 1D.
    Each physical state $x_i$ is split into two lifted states, typically interpreted as ``moving left'' or ``moving right''.
    Detailed balance is violated, but global balance is maintained: for every lifted state, the sums of incoming and outgoing equilibrium flows both equal the stationary probability.
    Unlike in the Metropolis scheme (Fig.~\ref{fig:metropolis}), there are no self-transitions and thus no rejections.}
    \label{fig:lifted}
\end{figure}

A simple example of a Markov chain that satisfies global balance but violates detailed balance is shown in Fig.~\ref{fig:lifted}.
Compared with the Metropolis scheme in Fig.~\ref{fig:metropolis}, each original state $x_i$ is split into two \emph{lifted} substates, labeled by an internal variable $v \in \{-1,+1\}$.
The total stationary probability of the pair of substates equals the original $\pi_i$, so that the marginal distribution over the physical state $x_i$ is unchanged.

The internal variable $v$ is naturally interpreted as a direction or velocity.
If a Metropolis proposal were rejected, the lifted scheme instead flips $v \to -v$.
Thus, transitions in the lifted state space consist of either ballistic motion at fixed $v$ or ``collisions'' that reverse $v$.
Although detailed balance is broken, the global balance condition is preserved because, for every lifted node $(v,x_i)$, the sum of incoming flows and the sum of outgoing flows both equal $\pi(v,x_i)$.
When attempting to move with a velocity $v$ from state $i$ to the next state $j=i+v$, the equilibrium probability flow for colliding and reversing the velocity is
\begin{equation}
\pi((v,x_i ); (-v,x_i)) = \left[ \frac{\pi_i}{2} - \frac{\pi_{i+v}}{2}\right]^+.
\end{equation}
This flow vanishes when attempting to move to a higher probability state ($\pi_{i+v}>\pi_{i}$); thus, collisions can only occur when attempting to move to a lower probability state.

At first sight, this may seem like a minor modification of Metropolis: rejection is merely replaced by flipping $v$.
However, the induced dynamics can be qualitatively different.
In the Metropolis chain, each node has an equal probability to move left or right, leading to diffusive dynamics.
In the lifted scheme, by contrast, collisions vanish in the interior: the system moves persistently in one direction as long as the probability of the equilibrium state increases (for the Boltzmann distribution, potential energy decreases) and might collide only when this is no longer the case.
The dynamics become convective rather than diffusive.
Crucially, the lifted scheme contains no self-arrows: in the lifted space, what would be a rejection in the Metropolis scheme is replaced by a switch of the internal state. It is strictly rejection-free.

Sometimes, it is claimed that a condition stronger than rejection-free, namely `maximal global balance', should be strived for \cite{Michel2014}. This condition requires that probability flows between any two states are unidirectional.
This prevents backflow and thus makes the flow maximally convective and minimally diffusive.
While this is a good guiding principle, it is not always achievable. For example, in Fig.~\ref{fig:lifted}, if the equilibrium probability $\pi_i$ has a local maximum (corresponding to a potential minimum), then the lifting must be bidirectional to satisfy global balance.
A second consideration is that such purely convective motion might adversely affect ergodicity. This becomes clear when investigating the limiting case of a uniform distribution $\pi_i = 1/N$.
A detailed analysis of mixing rates for a 1D chain, provided in Appendix~\ref{app:1D_mixing}, demonstrates that lifting can reduce the relaxation time scaling from $O(N^2)$ to $O(N)$ by introducing a small `spontaneous' collision probability.
In practice, minor ergodicity issues are sometimes encountered in ECMC schemes. These are often countered by resampling velocities in the system.
An alternative strategy would be introducing a small probability for spontaneous collisions, e.g.,
\begin{equation}
\pi((v,x_i ); (-v,x_i)) = \varepsilon \, \frac{\pi_i}{2} + (1-\varepsilon) \, \left[ \frac{\pi_i}{2} - \frac{\pi_{i+v}}{2}\right]^+.
\end{equation}

It is useful to generalize the 1D lifted Metropolis scheme to a setting where transport is ballistic, with intermittent changes in velocity due to collisions.
Assuming a Metropolis filter for the ballistic part, and writing the global-balance conditions for the flows [Eqs.~\eqref{eq:probability_flow} and~\eqref{eq:probability_flow_conditions}] in discrete time with step $\Delta t$, we obtain
\begin{align}
  \pi(v,x) &= \pi((v,x-v\Delta t); (v,x)) + \sum_{v'} \pi((v',x); (v,x)) \nonumber\\
           &= \min\!\big(\pi(v,x-v\Delta t), \pi(v,x)\big) + \sum_{v'} \pi((v',x); (v,x)), \\
  \pi(v,x) &= \pi((v,x); (v,x+v\Delta t)) + \sum_{v'} \pi((v,x); (v',x)) \nonumber\\
           &= \min\!\big(\pi(v,x), \pi(v,x+v\Delta t)\big) + \sum_{v'} \pi((v,x); (v',x)).
 \label{eq:lifted_Metropolis}
\end{align}

Our primary goal is to sample $\pi_x(x)$, the probability distribution over positions.
It is therefore convenient to assume that the distribution of $v$ is independent of $x$, and to introduce a conditional probability flow in velocity space,
\begin{equation}
\pi(v,x) = \pi_v(v)\,\pi_x(x),
\qquad
\pi(v; v'|x) \equiv \frac{\pi((v,x); (v',x))}{\pi_x(x)}.
\end{equation}
With these definitions, Eqs.~\eqref{eq:lifted_Metropolis} become
\begin{align}
  \pi_v(v)\,\pi_x(x)
    &= \pi_v(v)\,\pi_x(x)\,
       \min\!\left(1, \frac{\pi_x(x-v\Delta t)}{\pi_x(x)}\right)
       + \pi_x(x)\sum_{v'} \pi(v';v|x),
    \label{eq:lifted_global_balance}\\
  \pi_v(v)\,\pi_x(x)
    &= \pi_v(v)\,\pi_x(x)\,
       \min\!\left(1, \frac{\pi_x(x+v\Delta t)}{\pi_x(x)}\right)
       + \pi_x(x)\sum_{v'} \pi(v;v'|x).
    \label{eq:lifted_normalization}
\end{align}

The normalization relation in Eq.~\eqref{eq:lifted_normalization} is satisfied if we write
\begin{equation}
\pi(v; v'|x) = P_\mathrm{coll}(v,x)\,\pi_v(v)\,T_v(v\to v'|x),
\qquad
P_\mathrm{coll}(v,x)
= \left[1 - \frac{\pi_x(x+v\Delta t)}{\pi_x(x)} \right]^+,
\end{equation}
where $P_\mathrm{coll}(v,x)$ is the collision probability and $T_v(v\to v'|x)$ is the post-collision transition kernel in velocity space.
The global-balance condition, Eq.~\eqref{eq:lifted_global_balance}, then becomes
\begin{equation}
\left[1 - \frac{\pi_x(x-v\Delta t)}{\pi_x(x)}\right]^+ \pi_v(v)
= \sum_{v'} \left[1 - \frac{\pi_x(x+v'\Delta t)}{\pi_x(x)}\right]^+
  \pi_v(v')\,T_v(v'\to v|x).
\label{eq:requirement_collision_discrete}
\end{equation}
Except for special choices such as a deterministic collision rule $v'=-v$, it is generally hard to satisfy Eq.~\eqref{eq:requirement_collision_discrete} exactly for finite~$\Delta t$.

The situation simplifies in the limit $\Delta t\to 0$.
For a Boltzmann distribution [Eq.~\eqref{eq:Boltzmann_distribution}] one has
\begin{equation}
    \left[1 - \frac{\pi_x(x+v\Delta t)}{\pi_x(x)}\right]^+
    = \left[1 - \exp\big(-\beta\,[U(x+v\Delta t)-U(x)]\big)\right]^+
    \approx \beta\,[v\cdot\nabla U(x)]^+\,\Delta t,
\end{equation}
so Eq.~\eqref{eq:requirement_collision_discrete} becomes
\begin{equation}
[-v\cdot\nabla U(x)]^+\,\pi_v(v)
= \sum_{v'} [v'\cdot\nabla U(x)]^+\,\pi_v(v')\,T_v(v'\to v|x).
\label{eq:requirement_collision_continuous}
\end{equation}
This admits many possible solutions.
A simple and important class is characterized by
\begin{equation}
\sum_{v'} \pi_v(v')\,T_v(v'\to v|x) = \pi_v(v), \text{ with }
    T_v(v\to v'|x) = 0,\text{ if } v'\cdot\nabla U(x) \ne -v\cdot\nabla U(x),\label{eq:collsion_requirement}
\end{equation}

Thus, two key ingredients characterize this class of lifted schemes: first, collisions occur only when moving uphill in the potential ($v\cdot\nabla U>0$), and second, collisions reverse the component of the velocity along $\nabla U$.
These collisions must be constructed such that they leave a chosen velocity distribution, $\pi_v(v)$, invariant.
Since the primary goal is usually to sample the configurational distribution $\pi_x(x)$, the velocity $v$ acts as an auxiliary variable, offering some freedom in the choice of $\pi_v$.
However, a crucial constraint arises: while the collision transition probabilities $T_v(v'\to v|x)$ inherently depend on the position $x$ (via $\nabla U(x)$), the velocity distribution $\pi_v(v)$ must remain independent of $x$ to allow for the factorization $\pi(x,v)=\pi_x(x)\pi_v(v)$.
This requirement imposes a severe restriction on the allowed collision rules that satisfy Eq.~\eqref{eq:requirement_collision_continuous}.

\subsection{Factorized Metropolis filter}

The lifted Metropolis schemes discussed so far are conceptually useful but remain impractical for molecular simulations.
In particular, a collision in the lifted scheme typically reverses the velocity of \emph{all} particles involved in the configuration, which is far too global to be computationally efficient.
A crucial improvement is obtained by factorizing the Metropolis acceptance rate so that collisions become \emph{localized} to only those particles that contribute to a specific interaction.

In molecular simulations the total potential energy can almost always be expressed as a sum over interaction terms,
\begin{equation}
    U = \sum_{\alpha} U_{\alpha},
\end{equation}
where each $U_{\alpha}$ typically depends on only a few particles (e.g.\ pair interactions, angle terms, dihedral potentials).
As noted earlier (cf.\ Eq.~\eqref{eq:condition_minimum}), any acceptance rule that yields probability flows smaller than or equal to those of Metropolis still samples the correct equilibrium distribution.

For a proposed move with energy increments $\Delta U_{\alpha}$, the traditional Metropolis acceptance rule is
\begin{equation}
p_{\mathrm{acc}}
 = \min\!\left(1,\, \exp(-\beta \sum_{\alpha} \Delta U_{\alpha})\right)
 = \exp\!\left(-\beta \big[\sum_{\alpha} \Delta U_{\alpha}\big]^+\right).
\end{equation}
Using the inequality
\begin{equation}
[\sum_{\alpha} \Delta U_{\alpha}]^+
\,\le\,
\sum_{\alpha} [\Delta U_{\alpha}]^+,
\end{equation}
we obtain the \emph{factorized Metropolis filter} \cite{Michel2014}:
\begin{equation}
\min\!\left(1,\, e^{-\beta\sum_{\alpha} \Delta U_{\alpha}}\right)
\;\ge\;
\prod_{\alpha} e^{-\beta [\Delta U_{\alpha}]^+}.
\label{eq:factorized_Metropolis}
\end{equation}
In conventional Metropolis Monte Carlo, the right-hand side of Eq.~\eqref{eq:factorized_Metropolis} only increases the rejection rate and therefore provides no computational advantage.
In a lifted algorithm, however, this factorization is transformative.

In a lifted scheme, a Metropolis rejection becomes a \emph{collision}.
With the factorized filter, this collision probability is decomposed into independent contributions from each potential component $\alpha$.
Each factor in Eq.~\eqref{eq:factorized_Metropolis} then has a natural interpretation: $1-\exp\big(-\beta [\Delta U_{\alpha}]^+\big)$ is the probability that the system collides with potential term $\alpha$.

When a collision with interaction $\alpha$ occurs, the velocity update must reverse the component of $v$ along the local energy gradient (according to eq.~\eqref{eq:collsion_requirement}):
\begin{equation}
v'\cdot \nabla U_{\alpha} = -\, v\cdot \nabla U_{\alpha}.
\label{eq:valid_collision}
\end{equation}
Crucially, this constraint only involves the particles that appear in $U_{\alpha}$.
Thus, collisions are localized: only a small subset of particle velocities must be updated, typically a pair for pair potentials or a triplet for angle potentials.

The factorized Metropolis framework thus resolves a key limitation of lifted Metropolis schemes.
Instead of global velocity reversals that couple the entire system, collisions now correspond to localized interactions.
This serves as the conceptual bridge between lifted Markov chains and event-driven Monte Carlo algorithms such as the event-chain method and its generalizations: collisions occur independently for each interaction component, allowing the system to evolve deterministically between events.

\subsection{Event-driven implementation}

The collision requirements of the lifted Metropolis scheme, Eq.~\eqref{eq:requirement_collision_continuous}, are rigorously valid only in the limit of vanishing step size $\Delta t \to 0$.
To obtain a continuous-time description, we derive the transition probabilities from the equilibrium flows (Eq.~\eqref{eq:metropolis_transition}) and insert them into the master equation, Eq.~\eqref{eq:master}.
Taking the limit $\Delta t \to 0$ yields the partial differential equation
\begin{equation}
\frac{\partial p(x,v,t)}{\partial t}
= -\, v \cdot\!\nabla_x p(x,v,t)
  - \beta\, \sum_\alpha [v\cdot\!\nabla U_\alpha]^+ \, p(x,v,t)
  + \beta\, \sum_\alpha 
     \int_{v'} [v'\cdot\!\nabla U_\alpha]^+\, p(x,v',t)\, T_\alpha(v'\!\to v|x)\, dv'.
\end{equation}
The first term describes deterministic streaming with constant velocity $v$.
The second term represents the loss of probability due to collisions $v \to v'$, while the third term accounts for probability gained from collisions $v' \to v$.
As a check for consistency: For the equilibrium distribution, $p(x,v,t)=\pi_x(x) \, \pi_v(v)$ with $\pi_x$ the Boltzmann distribution, Eq.~\eqref{eq:Boltzmann_distribution}, one finds $-v \cdot\!\nabla_x \pi_x(x) \, \pi_v(v) = \beta v\cdot \nabla \sum_\alpha U_\alpha$. Inserting this relation, and using $a-[a]^+ = [-a]^+$, the requirement Eq.~\eqref{eq:requirement_collision_continuous} is recovered (with the sum replaced by an integral over velocities).

Over a small time interval $\Delta t$, the system either streams ballistically,
\begin{equation}
x(t+\Delta t) = x(t) + v\,\Delta t,
\end{equation}
or undergoes a collision that changes the velocity.
Let $t'$ be a later time.
The probability of \emph{no} collision having occurred up to $t'$ equals the product of the survival probabilities for each potential component $\alpha$:
\begin{equation}
P_{\mathrm{no\!-\!coll}}(t,t')
= \prod_{\alpha}
  \exp\!\left(
    -\beta \int_{t}^{t'} 
    \left[ \frac{dU_\alpha(t')}{dt'} \right]^+ dt'
  \right),
\qquad
U_\alpha(t') = U_\alpha\big(x(t) + v(t'-t)\big).
\end{equation}
Using inverse-transform sampling, the collision time associated with a specific potential term $\alpha$ is obtained by drawing a uniform random number $u\in (0,1)$ and solving
\begin{equation}
\int_{t}^{t_{\mathrm{coll},\alpha}}
\left[ \frac{dU_\alpha(t')}{dt'} \right]^+ dt'
= - \beta^{-1} \, \ln u.
\label{eq:collision_time}
\end{equation}
This yields a candidate collision time $t_{\mathrm{coll},\alpha}$ for each interaction $\alpha$.

The event-driven lifted Metropolis Monte Carlo method 
\cite{Peters2012,Michel2014,Kapfer2016} proceeds as follows:
\begin{enumerate}
\item For each potential component $\alpha$, compute the next candidate collision time $t_{\mathrm{coll},\alpha}$ by solving Eq.~\eqref{eq:collision_time}.
\item Identify the earliest collision time $t_{\mathrm{coll}} = \min_{\alpha} t_{\mathrm{coll},\alpha}$.
\item Move the system ballistically to this time.
\item Update the velocities of the particles participating in interaction $\alpha$ that caused the collision.
\item Recompute the collision times for all potential components involving the particles whose velocities were changed.
\item Repeat.
\end{enumerate}
For pair potentials, only two particle velocities are updated per collision; for angle or torsion terms, typically three or four particles are involved.
Thus, collisions remain strictly local despite the global nature of the full potential energy.

A practical implementation uses a priority queue (event heap) to store all candidate collision times.
To make this efficient for large molecular systems, several optimizations are typically employed.
Neighbor lists and cell lists are used to restrict the number of interaction terms $\alpha$ whose collisions need to be monitored, while bounding potentials provide inexpensive lower bounds on collision times so that costly evaluations are postponed until strictly necessary.
It is also advantageous to manage events in batches: collision times up to a chosen time horizon are inserted into the heap and maintained in sorted order, whereas events further in the future remain unsorted and are added only when required.
When the time horizon is reached, the heap is rebuilt from the next batch.
In combination, these strategies allow event-driven lifted Monte Carlo algorithms to maintain good performance.

\subsection{Collision detection and event scheduling}

The core computational task in event-driven MC is solving Eq.~\eqref{eq:collision_time} for the collision time $t_{\mathrm{coll}}$. While formally an integral, the problem reduces to finding the first time $t > t_{\mathrm{now}}$ where the accumulated positive change in potential energy equals a stochastic threshold $E^* = -\beta^{-1} \ln u$. 

For most standard force fields, an interaction potential $U_\alpha$ depends on a single scalar coordinate $q(t)$, such as a pair distance $r_{ij}$, a bond angle $\theta_{ijk}$, or a dihedral angle $\phi_{ijkl}$. The time derivative decomposes as $\dot{U}_\alpha = (dU_\alpha/dq) (dq/dt)$. Consequently, the integration domain is delimited by the roots of two functions: potential extrema ($dU_\alpha/dq = 0$) and trajectory extrema ($dq/dt = 0$).

The potential extrema are static properties of the potential function. The corresponding $q_n$ values can be precomputed and stored. Note that if the potential function is not differentiable, the discontinuities in $U_\alpha$ must also be tracked alongside the extrema. The trajectory extrema are geometric properties of the ballistic trajectory, such as the distance of closest approach for a pair or the turning point of an angle.

These extrema divide $U_\alpha(q)$ into segments, some of which do not contribute to the integral because $\mathrm{sgn}(dU_\alpha/dq) = -\mathrm{sgn}(dq/dt)$ (i.e., the potential energy is decreasing along the trajectory). Starting from the initial value $U_\alpha(q_0)$, one can form a cumulative sum of the energy changes over the intervals where $\dot{U}_\alpha(t) > 0$ and identify the specific interval $[q_n, q_{n+1}]$ where the accumulated energy change crosses the threshold $E^*$. For this interval, the target potential value at collision is given by:
\begin{equation}
    U_{\alpha,\mathrm{coll}} = U_\alpha(q_n) + E^* - \sum_{i=1}^{n} [U_\alpha(q_{i}) - U_\alpha(q_{i-1})]^+.
\end{equation}
It is possible that $E^*$ is too large and no such $U_{\alpha,\mathrm{coll}}$ exists within the valid trajectory range; this simply indicates that no collision occurs with $U_\alpha$ during the current move.
For complex potentials, the monotonic segments can be pre-processed and the inverse function $q(U)$ stored (e.g., via spline interpolation) to allow for rapid lookups of the collision coordinate, $q_{\mathrm{coll}} = q(U_{\alpha,\mathrm{coll}})$.

Finally, one solves the geometric equation $q(t_{\mathrm{coll}}) = q_{\mathrm{coll}}$ for $t_{\mathrm{coll}}$. This geometric problem can often be formulated as finding the roots of a low-degree polynomial. Note that finding the delimiting $q_n$ values that obey $dq/dt=0$ involves a similar root-finding problem.

A critical feature of event-driven algorithms is that any collision changes the velocity of the involved particles, instantaneously invalidating all other scheduled events involving them. In dense systems, a particle might be involved in hundreds of weak interactions, but only the earliest event is physically realized; computing exact collision times for all interactions is therefore computationally wasteful. Efficient implementations rely on three main strategies to minimize this overhead:

First, instead of solving the exact collision time immediately, one can compute a cheap lower bound $t_{\mathrm{bound}} \le t_{\mathrm{coll}}$. This lower bound is inserted into the event queue. When $t_{\mathrm{bound}}$ reaches the top of the queue, the exact time is computed. If the exact time is still the earliest, the event is executed; otherwise, the new exact time is re-inserted.

Second, for complex or long-range potentials where even the lower bound is expensive, one can use a simplified upper bound for the event rate, $\Lambda(t) \ge \lambda(t) = \beta[\dot{U}_\alpha(t)]^+$. Candidate events are generated efficiently using $\Lambda(t)$ and then accepted as real events with probability $\lambda(t) / \Lambda(t)$. Rejected candidates are simply discarded (``thinned''). This avoids the need to invert the potential.

Third, for weak, long-range interactions, the cumulative event rate is low. Instead of tracking each interaction individually, they can be grouped (e.g., all interactions between two spatial cells). A bounding rate $\Lambda_{\mathrm{group}}$ is estimated for the whole group. When a group event occurs, a specific interaction $\alpha$ is selected with probability proportional to its contribution. This is the basis of the \emph{Cell-Veto} algorithm~\cite{Kapfer2016}, which allows long-range forces to be handled with $O(1)$ complexity per event.

Finally, the maintenance of the event queue itself must be efficient. If the number of particles is not too large (say $N$ below $10^4$), a standard binary heap scaling as $O(\log N)$ is often sufficient, provided that efficient bookkeeping is implemented to quickly find and remove invalidated collisions. However, for highly optimized codes dealing with massive event rates, specialized structures like tournament trees (e.g., Paul's Queue) are often preferred to achieve $O(1)$ amortized performance.

\subsection{Collision handling}

A fundamental requirement for a valid collision is given by Eq.~\eqref{eq:valid_collision}, which dictates that the velocity component along the gradient $\nabla U_\alpha$ must change sign. This condition is naturally satisfied by a reflection. While global balance theoretically leaves the change in velocity components perpendicular to the gradient unconstrained, the simplest and most logical choice is to leave them unchanged.

The specific choice of reflection determines the form of the invariant velocity distribution, $\pi_v(v)$. To ensure compatibility with standard molecular dynamics (MD) codes and physical intuition, it is advantageous to target the Maxwell-Boltzmann distribution. This requires a collision rule that leaves the total kinetic energy, $K = \sum_i \frac{1}{2} m_i \mathbf{v}_i^2$, invariant. The appropriate reflection, acting on the velocities of all particles $i$ involved in interaction $\alpha$, is given by:
\begin{equation}
\mathbf{v}_i' = \mathbf{v}_i - 2 \frac{\sum_j \nabla_j U_\alpha \cdot \mathbf{v}_j}{\sum_j m_j^{-1} \,\nabla_j U_\alpha \cdot \nabla_j U_\alpha} \, m_i^{-1} \, \nabla_i U_\alpha.
\label{eq:maxwell_boltzmann_invariants}
\end{equation}

With this choice, the event-driven lifted Metropolis scheme preserves both the Maxwell-Boltzmann velocity distribution and the configurational part of the Boltzmann distribution. However, the resulting dynamics differ fundamentally from Newtonian mechanics. In standard NVE Molecular Dynamics, the total energy ($H = K + U$) is conserved, leading to a continuous interchange between kinetic and potential energy. In contrast, the collisional method defined here conserves kinetic energy exactly at every event. Consequently, there is no intrinsic mechanism for equilibration between kinetic and configurational degrees of freedom; the system samples the correct configurational distribution for the given inverse temperature $\beta$, but the aggregate velocity magnitude remains frozen at its initial value.

To define the general collision requirement formally, it is convenient to represent the full system velocity $v$ as a high-dimensional column vector $\mathbf{v}$. The requirement for collisions with potential contribution $\alpha$ is Eq.~\eqref{eq:collsion_requirement} specialized for $U_\alpha$. We introduce a projection operator $\mathbf{P}$ which, in matrix notation where $\nabla U_\alpha$ represents the vector of gradients for all particles, takes the form:
\begin{equation}
\mathbf{P} = \frac{\mathbf{M}^{-1} \, \nabla U_\alpha \,  \nabla U_\alpha^T \, }{\nabla U_\alpha^T \, \mathbf{M}^{-1} \, \nabla U_\alpha}.
\end{equation}
Here, $\mathbf{M}$ is a general positive-definite symmetric mass matrix.

A general scheme to generate a Gaussian velocity distribution via reflection can be constructed by mixing the deterministic reflection with a stochastic update for the orthogonal components:
\begin{equation}
\mathbf{v}' =  (\mathbf{I} - 2 \mathbf{P}) \, \mathbf{v} + (\mathbf{I}_\alpha  - \mathbf{P}) \, \left(-(1-a)  \, \mathbf{v} + \sqrt{1-a^2} \, \mathbf{v}_G \right), \quad \text{where } -1 \le a \le 1.
\label{eq:general_reflection_update}
\end{equation}
In this expression, the velocity component along the gradient is always reflected (the first term). The parameter $a$ tunes the behavior of the velocity components perpendicular to the gradient. $\mathbf{I}_\alpha$ denotes a diagonal matrix with entries of 1 for particles involved in interaction $\alpha$ and 0 otherwise, ensuring the update is local. $\mathbf{v}_G$ is a velocity vector drawn from the target Gaussian distribution.

The choice $a=1$ yields a deterministic collision that leaves the quadratic form $\mathbf{v}^T \mathbf{M} \mathbf{v}$ invariant; if $\mathbf{M}$ contains particle masses on its diagonal, Eq.~\eqref{eq:maxwell_boltzmann_invariants} is recovered. The choice $a=-1$ corresponds to a full velocity reversal for all particles in $\alpha$. Intermediate values mix in the stochastic velocity $\mathbf{v}_G$. Provided $\mathbf{v}_G$ is drawn from a distribution with covariance $\langle \mathbf{v}_G \, \mathbf{v}_G^T \rangle = \beta^{-1} \mathbf{M}^{-1}$, the invariant distribution $\pi_v$ for this generalized update is the Maxwell-Boltzmann distribution.

\subsection{Event-chain collisions}

The original Event-Chain Monte Carlo (ECMC) algorithm for hard spheres \cite{Bernard2009} can be understood as a specific, limiting case of the general event-driven lifted Metropolis scheme. Its defining characteristic is the restriction of the ``active'' state to a single particle at any given time.

This transfer of activity can be viewed as a special type of elastic collision where velocity is fully transferred from one particle to another. Consider a system where only particle $i$ is moving with velocity $\mathbf{v}$, while all other particles are stationary. The high-dimensional velocity vector of the system is $v = (0, \dots, 0, \mathbf{v}_i, 0, \dots, 0)$. A standard event-chain collision involves transferring this activity to a target particle $j$, such that in the post-collision state, $\mathbf{v}_i = 0$.

In the Straight Event-Chain (SEC) method, the velocity is transferred directly: $\mathbf{v}_j = \mathbf{v}_i = \mathbf{v}$. Other choices are possible; for instance, the Reflected Event-Chain (REC) method inverts the velocity components perpendicular to the potential gradient $\nabla U_\alpha$. For a pair potential, we have $\nabla_j U_\alpha = -\nabla_i U_\alpha$. Therefore, if the velocity component along the gradient is preserved during the transfer (i.e., $\mathbf{v}_i \cdot \nabla_i U_\alpha = \mathbf{v}_j \cdot \nabla_i U_\alpha$), the fundamental collision condition Eq.~\eqref{eq:valid_collision} is automatically obeyed.

In the limit of hard-sphere interactions, the ``collision'' becomes a hard constraint. There is no stochastic acceptance step; the event occurs precisely at contact. Because the scheme satisfies global balance, the canonical distribution is preserved without detailed balance. This allows for a unique, rejection-free mode of simulation: a purely deterministic chain. By fixing the total chain length (or time duration) and iterating through starting particles sequentially, the entire simulation becomes a deterministic map.

\begin{figure}[t]
\centering
\includegraphics[width=0.9\linewidth]{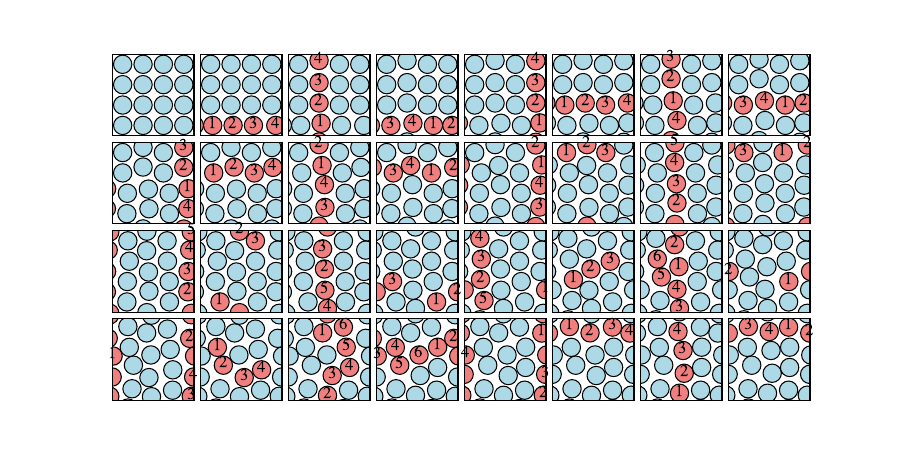}
\caption{Visualization of a purely deterministic ECMC trajectory in a 2D hard-disk system. The simulation is initialized on a lattice (box size $L=8$, disk radius $R=0.9$). The algorithm iterates through particles in sequence, moving them with a fixed event-chain length of $l=0.7\sqrt{2}$. The direction of motion alternates strictly between $+x$ and $+y$. The resulting ``woven'' pattern illustrates how the algorithm explores configuration space via cooperative, multi-particle chains rather than independent random walks, but still mixing the system.}
\label{fig:deterministic_hard_sphere_ecmc}
\end{figure}

Figure~\ref{fig:deterministic_hard_sphere_ecmc} visualizes this deterministic nature. It displays the traces of a 2D hard-disk system evolved using purely deterministic ECMC. The complex, weaving trajectories arise not from random kicks, but from the strict geometry of the particle packing and the sequential activation of collision partners. While the invariance of the equilibrium distribution is guaranteed by global balance, the irreducibility (ergodicity) of such a deterministic Markov chain is less obvious. A minimal requirement for mixing is that the set of lifting variables spans the space; for example, alternating between moves in the $x$ and $y$ directions as done in Fig.~\ref{fig:deterministic_hard_sphere_ecmc}. Theoretical guarantees are difficult to derive, but numerical experiments suggest that simple protocols—such as alternating directions or randomizing the chain length—are sufficient to ensure ergodicity in dense fluids.

To generalize the collision scheme beyond simple pair potentials, it is useful to shift perspective. Instead of viewing velocity as a quantity transferred between particles, we can describe the system such that \emph{all} particles possess a velocity, but only one, the \emph{active} particle, moves with it at any given time. All others remain frozen in place until activated.

With this shift in perspective, the single-particle ECMC scheme can be extended from pair potentials to general many-body potentials by expanding the state space description. We consider a state $(x,v,i)$, where $x$ and $v$ represent the positions and velocities of \emph{all} particles, and the index $i$ indicates the single currently active particle. The equilibrium distribution factorizes as:
\begin{equation}
\pi(x,v,i) = \pi_x(x) \, \pi_v(v) \, \frac{1}{N}.
\end{equation}
This assumes that, at equilibrium, every particle is equally likely to be the active one ($p=1/N$).

In this formalism, the active particle $i$ moves with velocity $\mathbf{v}_i$, creating a collision rate $[\mathbf{v}_i \cdot \nabla_i U_\alpha]^+$. When a collision with potential term $U_\alpha$ occurs, the algorithm may change the velocities of all participating particles and must select a new active particle. The generalized condition for collision transition probabilities is:
\begin{equation}
[-\mathbf{v}_i\cdot\nabla_i U_\alpha]^+\,\pi_v(v)
= \sum_{(v',j)} [\mathbf{v}_j'\cdot\nabla_j U_\alpha]^+\,\pi_v(v') \, T_\alpha((v',j)\to (v,i)|x).
\label{eq:requirement_collision_continuous_ECMC}
\end{equation}
Here, the sum is over all possible pre-collision states $(v', j)$ that could transition to the current state $(v, i)$.

A first solution to this equation is to require that all particles have the same fixed velocity vector $\mathbf{v}$ and to strictly transfer activity without changing velocities. Eq.~\eqref{eq:requirement_collision_continuous_ECMC} then simplifies to:
\begin{equation}
[-\mathbf{v}\cdot\nabla_i U_\alpha]^+
= \sum_{j} [\mathbf{v} \cdot\nabla_j U_\alpha]^+ \, T_\alpha(j\to i|x).
\label{eq:Harland_rule}
\end{equation}
Assuming the transition probability to particle $i$ depends only on the target state (and not the source $j$), we obtain the selection rule:
\begin{equation}
T_\alpha(j\to i|x) = \frac{[-\mathbf{v}\cdot\nabla_i U_\alpha]^+}{\sum_{k} [\mathbf{v} \cdot\nabla_k U_\alpha]^+ }.
\end{equation}
This rule effectively transfers activity to a particle $i$ that is ``pulling'' the system downhill (since $[-\mathbf{v}\cdot\nabla_i U_\alpha]^+ \ne 0$ only if the force opposes the velocity). We can verify that this is a valid probability distribution by summing over all possible next particles $i$. Using the identity $[-a]^+ = [a]^+ - a$ and the translational invariance of the potential ($\sum_i \nabla_i U_\alpha = 0$):
\begin{equation}
\sum_i T_\alpha(j\to i|x) = \frac{\sum_i [-\mathbf{v}\cdot\nabla_i U_\alpha]^+}{\sum_{k} [\mathbf{v} \cdot\nabla_k U_\alpha]^+ } = \frac{\sum_i [\mathbf{v}\cdot\nabla_i U_\alpha]^+ - \sum_i \mathbf{v}\cdot\nabla_i U_\alpha}{\sum_{k} [\mathbf{v} \cdot\nabla_k U_\alpha]^+ } = 1.
\end{equation}
This scheme generalizes the Straight Event-Chain (SEC) algorithm to translation-invariant multibody potentials and was introduced in Ref.~\cite{Harland2017} and applied to atomic force fields in Ref.~\cite{Faulkner2018}.

A more general solution allows for arbitrary particle velocities and velocity changes. We assume two possible outcomes for a collision: (1) velocities remain unchanged ($v \to v$) but activity transfers, or (2) velocities update ($v \to v'$) via a deterministic reflection (e.g.\  Eq.~\eqref{eq:maxwell_boltzmann_invariants}) and activity transfers. With $\pi_v(v)=\pi_v(v')$, the balance condition becomes:
\begin{equation}
[-\mathbf{v}_i\cdot\nabla_i U_\alpha]^+
= \sum_j \left( [\mathbf{v}_j \cdot\nabla_j U_\alpha]^+ \, T_\alpha((v,j) \to (v,i)|x) + [\mathbf{v}_j'\cdot\nabla_j U_\alpha]^+ \, T_\alpha((v',j)\to (v,i)|x) \right).
\label{eq:requirement_ECMC_2}
\end{equation}
If we again assume the transition probability depends only on the target state, we find the selection rule:
\begin{equation}
T_\alpha((v,j)\to (v,i)|x) = T_\alpha((v',j)\to (v,i)|x) = \frac{[-\mathbf{v}_i\cdot\nabla_i U_\alpha]^+}{\sum_k ( [\mathbf{v}_k \cdot\nabla_k U_\alpha]^+ \, + [\mathbf{v}_k'\cdot\nabla_k U_\alpha]^+)}.
\label{eq:selecion_rule}
\end{equation}
This defines a stochastic rule for handling an event triggered by active particle $j$. The algorithm considers all particles $k$ involved in $U_\alpha$ and two possible velocity states ($v$ and $v'$). It selects a new active particle $i$ and a velocity configuration with probability proportional to the rate at which that configuration moves downhill (relieving the potential energy).
Crucially, if the unkicked velocity state $v$ is chosen, the activity \emph{must} transfer to a new particle, because the current particle $j$ was moving uphill (triggering the event), so its weight $[-\mathbf{v}_j\cdot\nabla_j U_\alpha]^+$ is zero.

We verify normalization by summing over all target states (both $v$ and $v'$):
\begin{equation}
  \begin{split}
\sum_i \left( T_\alpha((v,j)\to (v,i)) + T_\alpha((v,j)\to (v',i)) \right)
&= \frac{\sum_i ( [-\mathbf{v}_i \cdot\nabla_i U_\alpha]^+ \, + [-\mathbf{v}_i'\cdot\nabla_i U_\alpha]^+)}{\sum_k ( [\mathbf{v}_k \cdot\nabla_k U_\alpha]^+ \, + [\mathbf{v}_k'\cdot\nabla_k U_\alpha]^+)} \\
&= 1 - \frac{\sum_i (\mathbf{v}_i \cdot\nabla_i U_\alpha + \mathbf{v}_i'\cdot\nabla_i U_\alpha)}{\sum_k ( [\mathbf{v}_k \cdot\nabla_k U_\alpha]^+ \, + [\mathbf{v}_k'\cdot\nabla_k U_\alpha]^+)} = 1.
  \end{split}
\end{equation}
The extra term vanishes because the reflection rule (Eq.~\ref{eq:valid_collision}).
With reflections according to Eq.~\eqref{eq:maxwell_boltzmann_invariants}, the algorithm realizes elastic collisions that conserve total kinetic energy (and linear and angular momentum) while maintaining the rejection-free, global-balance characteristics of the event chain. This approach was originally introduced for hard-sphere simulations as ``Newtonian'' ECMC \cite{Klement2019}, and is referred to as Generalized Newtonian ECMC in the context of many-body potentials \cite{Hollmer2024}.

\section{Discussion}

The development of Event-Chain Monte Carlo (ECMC) and its generalizations marks a paradigm shift in our understanding of sampling algorithms. For decades, detailed balance was viewed as a necessary evil: a strict symmetry condition imposed to guarantee correctness, at the cost of diffusive random-walk dynamics ($x \sim \sqrt{t}$). The seminal work of Bernard et al.~\cite{Bernard2009}, and its subsequent generalizations~\cite{Peters2012, Michel2014}, demonstrate that this constraint is artificial. By breaking detailed balance while enforcing the more fundamental condition of global balance, one can construct non-reversible, ballistic Markov chains ($x \sim t$) that converge to equilibrium orders of magnitude faster.

The development of the field following this breakthrough reveals a clear intellectual progression. Bernard et al.\ provided the physical proof-of-concept with hard spheres, demonstrating that "lifting" a rejection into a redirection creates persistent probability flows. Peters and de~With~\cite{Peters2012} generalized this to continuous potentials by introducing the concept of stochastic event rates, effectively bridging the gap between Monte Carlo and event-driven Molecular Dynamics. Finally, Michel, Kapfer, and Krauth~\cite{Michel2014} provided the rigorous unification, showing how the factorized Metropolis filter allows these continuous-time processes to be treated as mathematically exact infinitesimal steps.

This efficiency has enabled scientific breakthroughs that were previously computationally intractable. The most prominent example is the resolution of the melting transition in two-dimensional hard disks. The massive sampling power of ECMC allowed Bernard and Krauth to conclusively identify the existence of the hexatic phase, settling a decades-old controversy~\cite{Bernard2011}. Beyond hard spheres, the method has found success in dense polymer melts, where event chains facilitate reptation-like moves without steric rejection~\cite{Kampmann2015}. More recently, Castagn\`{e}de et al.\ (2024) employed ECMC to access the nucleation of Laves phases in soft-sphere systems at extremely low temperatures—a regime where standard Molecular Dynamics stalls due to the vanishingly small time steps required for steep repulsive potentials~\cite{Castagnde2024}. In spin systems, the method has been shown to effectively eliminate critical slowing down~\cite{Michel2015, Nishikawa2015}.

Crucially, the impact of these ideas extends well beyond physics. The EDMC formalism introduced by Peters and de~With~\cite{Peters2012} was adapted by the statistics community into the ``Bouncy Particle Sampler'' (BPS)~\cite{Bouchard-Ct2018}. This catalyzed the rapid development of Piecewise Deterministic Markov Processes (PDMP) for Bayesian inference, providing a powerful new tool for sampling high-dimensional posterior distributions in machine learning~\cite{Bierkens2019}. This cross-disciplinary adoption validates the fundamental nature of the event-chain insight.

Beyond these immediate applications, the generalized framework reviewed here laid the groundwork for two major subsequent algorithmic breakthroughs that addressed long-standing bottlenecks in the field.

First, the handling of long-range interactions, traditionally a cubic-scaling nightmare for event-driven methods, was solved by the Cell-Veto Monte Carlo algorithm~\cite{Kapfer2016}. By combining the factorized Metropolis filter with bounding potentials, this method achieves $O(1)$ complexity per particle move for long-range Coulombic systems without approximations, cutoffs, or Ewald summation grids. This effectively removed the computational ceiling for applying ECMC to charged atomistic systems.

Second, the empirical speedup of the method has recently found a rigorous analytical foundation. While early justifications relied on heuristic comparisons of dynamical exponents ($z$), recent solutions of the Lifted TASEP (Totally Asymmetric Simple Exclusion Process)~\cite{Essler2024, Kapfer2017} have proven mathematically that non-reversible, lifted chains belong to a faster universality class than their reversible counterparts. These analytic results confirm a polynomial reduction in relaxation times (e.g., from $\sim N^{5/2}$ to $\sim N^2$ or better), providing the theoretical justification for the efficiency observed in complex simulations.

However, despite these theoretical and algorithmic triumphs, a practical computational bottleneck remains. Modern high-performance computing is dominated by GPUs and massive parallelism. While standard Molecular Dynamics (MD) scales effortlessly via domain decomposition, standard event-driven MC is inherently sequential. This serial nature is efficient for mixing (algorithmic speedup) but limits the wall-clock throughput on hardware designed for millions of simultaneous threads.

\section{Outlook}

The application of event-driven MC lies in resolving the tension between its superior algorithmic scaling (mixing rate) and the hardware efficiency (FLOPS) of modern architectures. For simple fluids, highly optimized parallel MD will likely remain dominant. However, for glassy, frustrated, or extremely dense systems, raw FLOPS are insufficient if the system remains trapped in local basins. Here, the rejection-free nature of event-driven MC provides a decisive advantage.

We are already seeing the transition of these methods from ``toy models'' to full-scale molecular simulation. Recent work has demonstrated approximation-free sampling of dense all-atom systems, such as the SPC/Fw water model~\cite{Hollmer2024, Faulkner2018}. Unlike standard MD, which requires thermostats and time-step integration errors, these modern event-driven MC implementations sample the canonical ensemble exactly. Furthermore, variants like Newtonian Event Chains~\cite{Klement2019, Klement2021} and Irreversible Swap algorithms~\cite{Nishikawa2025, Ghimenti2024} have been developed to further accelerate dynamics in high-density regimes.

The event-driven nature makes massive parallelization difficult. One approach is to handle events in parallel using multiple threads. If the number of threads is much less than the number of particles, one can run multiple event chains in parallel with rare interference. This strategy has been implemented for event-chain simulation of hard spheres~\cite{Li2021}. Their approach handles conflicts by keeping the number of threads $k$ low enough ($k \propto \sqrt{N}$) that collisions between active chains are statistically rare. If a conflict is detected (a horizon violation), the update is aborted and restarted from the last valid breakpoint using new randomly selected active particles. While the rigorous validity of this abort-and-restart procedure is complex to prove, it appears robust when conflicts are rare.

An alternative implementation could involve generating tentative chains in parallel from an initial configuration without immediate updates. Once all chains are generated, involved particles are flagged. Particles flagged multiple times constitute a conflict that are resolved by determining priority. Valid segments of chains are kept, while conflicting portions are discarded and regenerated until all conflicts are resolved. This procedure would converge quickly for sparse conflicts and yield linear speedup with the number of threads, provided the thread count remains small relative to $N$. This method is well-suited for multithreading on CPUs, but the limited parallelism and complex conflict resolution make it less suitable for GPU implementation.

In the context of Event-Driven Molecular Dynamics (EDMD) and its rejection-free MC generalizations, parallelization is notoriously problematic. Research literature has investigated spatial decomposition with optimistic synchronization for EDMD~\cite{Miller2004}. However, these methods have not been widely adopted because their performance gains rarely exceed those of highly optimized serial implementations like the DynamO code~\cite{Bannerman2011}.

A second parallelization approach is domain decomposition with ``frozen'' checkerboard configurations~\cite{Anderson2013}. This has been successfully applied to Metropolis MC, where moves into frozen regions are simply rejected to maintain detailed balance. For ECMC, this is adapted by reflecting off the domain boundaries instead of rejecting~\cite{Kampmann2015_2}.
However, the freezing profoundly affects the dynamics. Metropolis MC is inherently diffusive, so using many small domains with frequent unfreezing is effective and maps well to massive GPU parallelism. In contrast, event-driven MC relies on ballistic motion over correlation lengths to achieve equilibration. When domains become too small, this ballistic length scale is truncated, and the effective motion reverts to diffusion, degrading the algorithmic advantage. Consequently, domain decomposition for EDMC appears viable only with large domains on CPU clusters (hybrid OpenMP/MPI) and seems unfavorable for massive GPU adoption, except perhaps for high-density systems where the rejection-free benefit outweighs the loss of ballistic transport.

Most research on event-driven MC focuses on single-particle ECMC. The generalization to molecular systems is elegant, and the cell-veto method for long-range forces is highly innovative. However, performance benchmarks~\cite{Hollmer2024} indicate that execution time is dominated by these long-range forces. Despite its conceptual beauty, making ECMC competitive with high-performance MD codes for general atomistic simulations remains a significant challenge.

A second, largely unexplored avenue is the hybridization of event-driven MC with Molecular Dynamics. This would utilize the all-particle variant~\cite{Peters2012} where all particles move simultaneously, as in hard-sphere EDMD. Although this dynamics differs from Newtonian mechanics, it possesses well-defined velocities and preserves the Boltzmann distribution. We envision an operator-splitting future where standard MD handles long-range forces while EDMC handles local interactions. In such a scheme, velocities would be updated by long-range MD forces, followed by an EDMC propagation step for bonded forces, and so on. By applying EDMC exclusively to bonded interactions, individual molecules could be handled in parallel (e.g., multithreaded). Since bonded interactions are stiff, this would replace ad-hoc constraints (like SHAKE) or alleviate time-step restrictions with statistically exact, rejection-free MC. Such a hybrid approach could also be powerful for backmapping coarse-grained structures, where MD often fails due to overlaps, but EDMC naturally resolves conflicts via collisions.

In conclusion, the legacy of the work of Bernard, Krauth, and Wilson is the realization that the ``rejection'' in Monte Carlo is not a failure, but an opportunity. By lifting rejections into new trajectories, EDMC turns the energy barrier into an engine of motion. Event-chain MC has significantly contributed to the simulation of dense systems with hard interactions. While transforming EDMC into a generic competitor for massive atomistic simulations seems unlikely due to parallelization hurdles, there remain significant unexplored application areas where it could have a transformative impact.

\appendix
\section{Mixing rates in 1D lifted chains}
\label{app:1D_mixing}

To illustrate the qualitative difference between diffusive Metropolis dynamics and convective lifted dynamics, consider a simple 1D chain with uniform distribution $\pi_i = 1/N$.
Purely convective motion lacks mixing and is not irreducible: the chain decomposes into disconnected cycles.
A simple remedy for this 1D case is to introduce a small collision probability $\varepsilon$.
The corresponding equilibrium flows can be written as
\begin{align}
\pi((v,x_i); (v,x_{i+v})) &= (1-\varepsilon)\,
   \min\!\left(\pi(v,x_i),\, \pi(v,x_{i+v})\right), \qquad v=\pm1, \\
\pi((v,x_i); (-v,x_i)) &= \varepsilon\, \pi(v,x_i)
  + (1-\varepsilon)\, [\pi(v,x_i) - \pi(v,x_{i+v})]^+ ,
  \label{eq:lifted_metropolis_1D_appendix}
\end{align}
where $[a]^+ \equiv \max(0,a)$.

For a uniform distribution $\pi(v,x_i)=1/(2N)$ these reduce to the particularly simple transition probabilities
\begin{equation}
T((v,x_i)\!\to\!(v,x_{i+v})) = 1-\varepsilon,\qquad
T((v,x_i)\!\to\!(-v,x_i)) = \varepsilon,
\label{eq:transition_lifted_uniform_appendix}
\end{equation}
which are essentially the dynamics introduced by Diaconis, Holmes, and Neal~\cite{Diaconis2000}.
At the boundaries of a finite chain the probability to flip direction is then set to~1.

To compare mixing rates, consider a periodic chain with $N$ nodes.
Because of translational invariance, eigenmodes can be written as
\begin{equation}
p^n(v,x_j) = z^n a_v \,\exp\!\left(i k x_j\right),
\end{equation}
with $k$ an integer multiple of $2\pi/L$ and $x_j = j (L/N)$.
Inserting this Ansatz into the master equation and using the transition probabilities in Eq.~\eqref{eq:transition_lifted_uniform_appendix} leads to an eigenvalue
\begin{equation}
z = (1-\varepsilon)\,\cos(k\Delta x)
    + i\, \sqrt{(1-2\varepsilon) - (1-\varepsilon)^2\cos^2(k\Delta x)}.
\end{equation}
For fast convergence, $|z|$ should be as small as possible.
When the square root is imaginary, $|z| = 1-2\varepsilon$, suggesting that large $\varepsilon$ is beneficial.
However, as soon as the argument of the square root becomes positive, the eigenvalue acquires a real part and $|z|$ increases again.
The optimal $\varepsilon$ occurs when the argument of the square root vanishes at the smallest nonzero wavenumber $k\Delta x = 2\pi/N$, yielding
\begin{equation}
\varepsilon
= \frac{\sqrt{2}\,\sin(\pi/N)
   + \cos(2\pi/N) - 1}{\cos(2\pi/N)}
\approx \frac{\sqrt{2}\,\pi}{N},
\end{equation}
and therefore
\begin{equation}
|z|^n \approx \exp\!\left(-\frac{2\sqrt{2}\pi\, n}{N}\right).
\end{equation}
The relaxation time thus scales linearly with $N$.
This should be contrasted with the standard Metropolis scheme for a uniform distribution, where $T(x_i \to x_{i\pm1}) = 1/2$ and the slowest Fourier mode satisfies
\begin{equation}
z = \cos\!\left(\frac{2\pi}{N}\right)
\approx 1 - \frac{2\pi^2}{N^2},
\qquad
|z|^n \approx \exp\!\left(-\frac{2\pi^2 n}{N^2}\right),
\end{equation}
which exhibits the familiar diffusive relaxation time that scales as $N^2$.

A further instructive test case is a monotonic distribution with $\pi_{i+1} < \pi_i$ for all $i$, corresponding to a strictly increasing potential.
In this situation, moves in the uphill direction have a finite collision probability, while downhill moves do not.
As a result, once the lifted chain flips to the downhill branch, it travels deterministically back towards low potential.
This strong directional asymmetry is characteristic of lifted chains that violate detailed balance.

\printbibliography
\end{document}

%% file: config.tex
\usepackage{bm}
\usepackage{amsmath}
\usepackage{amssymb}
\usepackage{epstopdf} 
\usepackage{wrapfig}  
\usepackage{url}
\usepackage{xpatch}
\usepackage{dcolumn}
\newcolumntype{d}[1]{D{.}{.}{#1}} 

\usepackage[style=numeric-comp,sorting=none,backend=biber, maxbibnames=50,giveninits=true]{biblatex}

\usepackage{hyperref}
\usepackage{xurl}
\hypersetup{breaklinks=true}
\renewbibmacro{in:}{}
\DeclareBibliographyAlias{article}{custom}
\DeclareFieldFormat{pages}{#1}
\DeclareFieldFormat{url}{\url{#1}}
\DeclareBibliographyDriver{article}{%
  \usebibmacro{bibindex}%
  \usebibmacro{begentry}%
  \usebibmacro{author/editor+others/translator+others}%
  \setunit{\labelnamepunct}\newblock
  \usebibmacro{title}%
  \newunit\newblock
  \usebibmacro{byauthor}%
  \newunit\newblock
  \usebibmacro{byeditor+others}%
  \newunit\newblock
  \printfield{version}%
  \newunit\newblock
  \printfield{journaltitle}%
  \newunit\newblock
  \iffieldundef{volume}
    {}
    {\printfield{volume}}
  \iffieldundef{number}
    {}
    {\printtext{(\printfield{number})}}
  \iffieldundef{pages}
    {}
    {\printtext{:\printfield{pages}},}
  \newunit\newblock
  \printfield{year}
  \newunit\newblock
  \iftoggle{bbx:eprint}
    {\usebibmacro{eprint}}
    {}%
  \iffieldundef{url}
    {}
    {\printtext{\printfield{url}}}
  \usebibmacro{finentry}%
}

\DeclareBibliographyDriver{misc}{%
  \usebibmacro{bibindex}%
  \usebibmacro{begentry}%
  \usebibmacro{author/editor+others/translator+others}%
  \setunit{\labelnamepunct}\newblock
  \usebibmacro{title}%
  \newunit\newblock
  \usebibmacro{byauthor}%
  \newunit\newblock
  \printfield{version}%
  \newunit\newblock
  \newunit\newblock
  \printfield{journaltitle}%
  \newunit\newblock
  \iffieldundef{volume}
    {}
    {\printfield{volume}}
  \iffieldundef{number}
    {}
    {\printtext{(\printfield{number})}}
  \newunit\newblock
  \iffieldundef{pages}
    {}
    {\printtext{:\printfield{pages},}}
  \newunit\newblock
  \newunit\newblock
  \iftoggle{bbx:eprint}
    {\usebibmacro{eprint}}
    {}%
  \newunit\newblock
  \usebibmacro{url+urldate}%
  \newunit\newblock%
  \usebibmacro{addendum+pubstate}%
  \setunit{\bibpagerefpunct}\newblock
  \usebibmacro{pageref}%
  \newunit\newblock
  \iftoggle{bbx:related}
    {\usebibmacro{related:init}%
     \usebibmacro{related}}
    {}%
  \usebibmacro{finentry}%
}

\DeclareFieldFormat{shorthandwidth}{#1}
\DeclareRangeChars{-}
\DeclareRangeCommands{\bibrangedash}

\DeclareCiteCommand{\cite}[\mkbibbrackets]
  {\usebibmacro{cite:init}%
   \usebibmacro{prenote}}
  {\usebibmacro{citeindex}%
   \usebibmacro{cite:comp}}
  {}
  {\usebibmacro{cite:dump}%
   \usebibmacro{postnote}}